\documentclass[onecolumn,aps,pra,reprint,superscriptaddress,longbibliography]{revtex4-1}
\usepackage{amsmath}
\usepackage{graphicx}
\usepackage[lofdepth,lotdepth, caption=false]{subfig}
\usepackage{verbatim}
\usepackage{color}
\usepackage{dcolumn}
\usepackage{bm}
\usepackage{miller}
\usepackage{color}
\usepackage{adjustbox}

\usepackage{amsfonts}

\usepackage{setspace}

\usepackage{xr-hyper}
\usepackage{hyperref}

\usepackage[english]{babel}

\newcommand{\beginsupplement}{%
        \setcounter{table}{0}
        \renewcommand{\thetable}{S\arabic{table}}%
        \setcounter{figure}{0}
        \renewcommand{\thefigure}{S\arabic{figure}}%
     }

\begin{document}

\title{Observation of helimagnetism in the candidate ferroelectric CrI$_2$}

\author{John A.~Schneeloch}
\affiliation{Department of Physics, University of Virginia, Charlottesville,
Virginia 22904, USA}
%\email{jas9db@virginia.edu}

\author{Shunshun Liu}
\affiliation{Department of Materials Science and Engineering, University of Virginia, Charlottesville,
Virginia 22904, USA}

\author{Prasanna V. Balachandran}
\affiliation{Department of Materials Science and Engineering, University of Virginia, Charlottesville,
Virginia 22904, USA}
\affiliation{Department of Mechanical and Aerospace Engineering, University of Virginia, Charlottesville,
Virginia 22904, USA}

\author{Qiang Zhang}
\affiliation{Neutron Scattering Division, Oak Ridge National Laboratory, Oak Ridge, Tennessee 37831, USA}

\author{Despina Louca}
\thanks{Corresponding author}
\email{louca@virginia.edu}
\affiliation{Department of Physics, University of Virginia, Charlottesville,
Virginia 22904, USA}

\begin{abstract}
CrI$_{2}$ is a quasi-one dimensional (1D) van der Waals (vdW) system that exhibits helimagnetism that propagates along the ribbons. This was determined from neutron time-of-flight diffraction measurements.
Below $T_N=17$ K, a screw-like helimagnetic order develops with an incommensurate wavevector of $\mathbf{q} \approx (0.2492,0,0)$ at 8 K. Using density functional theory (DFT)$+U$ calculations, the $J_{1}$-$J_{2}$ model was leveraged to describe the helimagnetism, where $J_{1} (> 0)$ and $J_2 (< 0)$ correspond, respectively, to a ferromagnetic nearest neighbor (NN) and antiferromagnetic next-nearest neighbor (NNN) intrachain interaction. The DFT$+U$ calculations predict that bulk CrI$_2$ in the orthorhombic $Cmc2_1$ crystal structure satisfies the $|J_2| > |J_1|/4$ condition, which favors formation of helimagnetic order.
 
\end{abstract}

\maketitle

\pagestyle{plain}

\section{Introduction}
Van der Waals (vdW) layered materials have a broad range of potential applications, especially when magnetic order can be manipulated \cite{mcguireCleavableMagneticMaterials2020}. Efforts in this direction surged after reports of ferromagnetic (FM) order in monolayer CrI$_3$ \cite{huangLayerdependentFerromagnetismVan2017}, antiferromagnetic (AFM) order in monolayer FePS$_3$ \cite{leeIsingTypeMagneticOrdering2016, wangRamanSpectroscopyAtomically2016}, and the discovery that CrI$_3$ can be AFM or FM depending on layer stacking \cite{huangEmergentPhenomenaProximity2020,schneelochAntiferromagneticferromagneticHomostructuresDirac2023}. One vdW-layered magnetic material, CrI$_2$, has received little attention even thought its quasi-one dimensional (1D) structure offers a new view into the nature of exchange interactions in vdW heterostructures.

\begin{figure}[h!]
\begin{center}
\includegraphics[width=8.6cm]
{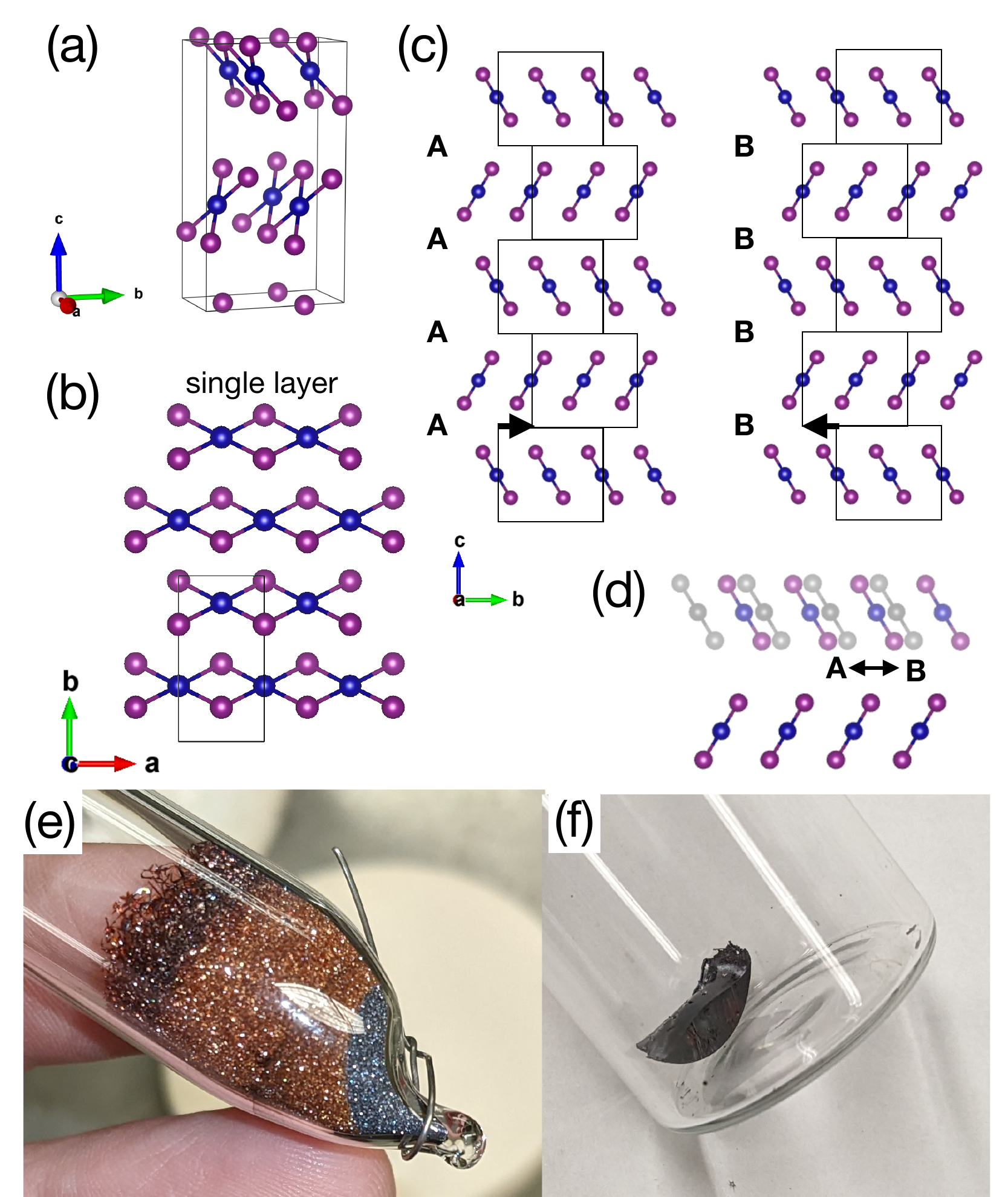}
\end{center}
\caption{(a) The crystal structure of orthorhombic CrI$_2$, showing ribbon chains along the $a$-axis. (b) One layer of CrI$_2$ in the $ab$ plane. (c) Both twins of orthorhombic CrI$_2$ depicted in the $bc$ plane, constructed by A- and B-type stacking, respectively. (d) An illustration that A- and B-type stacking are related by translations along the $b$-axis.
(e) Photo of a CrI$_2$ synthesis attempt, in which the ampoule (with stoichiometric Cr and I) was hung in a vertical tube furnace at 600 $^{\circ}$C. The boundary between the silvery  (CrI$_3$) and red (CrI$_2$) powder suggests the influence of a temperature gradient, where CrI$_3$ formed in a slightly cooler region of the ampoule. (f) Photo of a 0.5 g melt-grown piece of CrI$_2$. Structures plotted using \textsc{VESTA} \cite{mommaVESTAThreedimensionalVisualization2011}.}
\label{fig:1}
\end{figure}

CrI$_2$ has two polytypes with the same intralayer structure: a monoclinic phase, where the layers are oriented in the same direction \cite{tracyCrystalStructureChromium1962}, and an orthorhombic phase, where the layers are oriented in alternating directions 
\cite{guenEtudeSystemeChromeiode1972,besrestStructureCristallineIodure1973,guenSystemesCrI2MI2Ti1976,guenManganeseIIChrome1976}, the latter being reminiscent of the $T_d$ phase of MoTe$_2$ or WTe$_2$ \cite{clarkeLowtemperatureStructuralPhase1978}. The intralayer structure consists of a triangular lattice of Cr$^{2+}$ ions, each surrounded by six I$^-$ ions in an octahedral configuration, but with a Jahn-Teller (JT) distortion \cite{besrestStructureCristallineIodure1973} lengthening two of these bonds so that edge-sharing octahedra form ``ribbon chains'' along the $a$-axis (Fig.\ \ref{fig:1}(a,b)). The JT distortion induces a similar ribbon-chain structure in many other compounds, including CuBr$_2$ \cite{oecklerRedeterminationCrystalStructure2000} and CuCl$_2$ \cite{wells333CrystalStructure1947}, which are isostructural to monoclinic CrI$_2$, as well as CrBr$_2$ (which has a different layer stacking) \cite{tracyCrystalStructureChromium1962a}, CrCl$_2$ (in which the ribbon chains do not form CrI$_2$-like layers) \cite{tracyCrystalStructureChromium1961} and $\alpha$-PdCl$_2$ \cite{wellsCrystalStructurePalladous1939,eversStructuresDPdCl2GPdCl22010}.

The magnetic properties of CrI$_2$ are largely unexplored, with the sole attempt, to our knowledge, being a M\"{o}ssbauer spectroscopy study indicating a magnetic splitting of spectral lines at 4.2 K, but not at 77 K \cite{djermouniStructuralMagneticProperties1976}; a transition at 45(2) K was claimed, but without supporting data, and magnetic susceptibility was attempted but deemed inconclusive due to suspected contamination from the ferromagnetic CrI$_3$. 
Despite the paucity of bulk experimental studies of CrI$_2$, there have been three recent studies on monolayers synthesized by molecular beam epitaxy \cite{caiMolecularBeamEpitaxy2021,pengMottPhaseVan2020,liTwoDimensionalMagneticSemiconducting2023}, with scanning tunneling spectroscopy showing band gaps of about 3.0 to 3.2 eV \cite{caiMolecularBeamEpitaxy2021,pengMottPhaseVan2020}. Several recent theoretical studies have also been reported \cite{kulishSinglelayerMetalHalides2017,zhangAntiferromagneticTwodimensionalMaterial2020,zhaoTransitionCrI2Twodimensional2021,yangInterfacialTriferroicityMonolayer2022,zhangStructuralPhaseTransitions2022}, focusing mainly on the monolayer. In calculations, both ferromagnetic (FM) \cite{pengMottPhaseVan2020} and antiferromagnetic (AFM) \cite{zhangStructuralPhaseTransitions2022} orders result in the lowest free energy, though incommensurate ordering was not considered. 
Recent reports suggested that the orthorhombic phase has an out-of-plane polarization that can be reversed by sliding the layers \cite{zhangStructuralPhaseTransitions2022}, where such multiferroic behavior could lead to significant technological applications.
Thus, there is a clear need for experimental measurements to clarify the magnetic structure of this compound.

We report time-of-flight neutron diffraction measurements on a predominantly orthorhombic CrI$_2$ powder sample. Helimagnetic order is observed with a modulation wavevector $\mathbf{q}=(\delta,0,0)$ arising below $T_N=17$ K with $\delta \approx 0.2492$ at 8 K and decreasing slightly on warming. DFT$+U$ calculations were performed to extract the exchange interactions, resulting in a sizable next-nearest-neighbor AFM intrachain coupling that suggests an origin for the helimagnetic ordering.

\section{Methods}
\subsection{Experimental Details}
CrI$_2$ powder was synthesized by sealing a stoichiometric mixture of fine (-200 mesh) chromium powder and iodine pieces in ampoules and heating to 650 $^{\circ}$C within a day, then heated at 650 $^{\circ}$C or 750 $^{\circ}$C for several days before furnace-cooling. Below $\sim$600 $^{\circ}$C, CrI$_3$ forms instead, and an ampoule heated at 600 $^{\circ}$C in a vertical tube furnace showed a sharp divide between silvery CrI$_3$ and red CrI$_2$ powder (Fig.\ \ref{fig:1}(e)), due to CrI$_3$ forming in the cooler region of the temperature gradient. For the single crystals used for the magnetic susceptibility, CrI$_2$ crystals were grown from a melt by heating an ampoule with a stoichiometric mixture of Cr and I up to 900 $^{\circ}$C (after dwelling at 650 $^{\circ}$C for a day to make sure the iodine had fully reacted with the chromium powder), then furnace-cooled to 750 $^{\circ}$C before quenching in water. Slow cooling from 900 $^{\circ}$C to 750 $^{\circ}$C resulted in larger crystals, such as the 0.5 g piece shown in Fig.\ \ref{fig:1}(f). CrI$_2$ is hygroscopic, and crystals disintegrate within minutes in humid air, though they can be stored in dry air for weeks without visible deterioration. Neutron diffraction experiments were performed on the POWGEN diffractometer at the Spallation Neutron Source of Oak Ridge National Laboratory \cite{huqPOWGENRebuildThirdgeneration2019}. The powder was loaded into a vanadium can in a helium glovebox before measurement. A POWGEN automatic changer was adopted as the sample environment. Two neutron banks with center wavelengths of 1.5 \AA\ and 2.665 \AA\ were used for the data collection. 
Refinement was done using calculations in which the diffuse scattering was obtained from the formalism of Ref.\ \cite{treacyGeneralRecursionMethod1991}; the software \textsc{GSAS-II} \cite{tobyGSASIIGenesisModern2013} was used to obtain lattice parameters. 
Single crystal X-ray measurements were also carried out as a function of temperature, with data refined using the software \textsc{APEX4} by Bruker. 

\subsection{Density functional theory}
First-principles calculations were performed based on density functional theory (DFT) as implemented in the open-source plane-wave pseudopotential code, \textsc{Quantum ESPRESSO} \cite{QE-2009, QE-2017}. We chose the generalized gradient approximation (GGA) and PBEsol exchange-correlation functional for all ground-state calculations \cite{Perdew1996, Perdew2008}. A DFT$+U$ approach with a Hubbard $U = 3$~eV \cite{zhaoTransitionCrI2Twodimensional2021} was applied on the Cr-$3d$ orbitals to account for the correlated orbital effects \cite{HubbardU1991,HubbardU1994}. The kinetic energy cutoff was set to 70~Ry for the wavefunctions, and the Brillouin zone was sampled using a $\Gamma$-centered Monkhorst-Pack $k$-mesh for all calculations \cite{Monkhorst_Pack}. Semiempirical Grimme DFT-D3 vdW dispersion correction approach with three-body contribution was also applied to treat the vdW bonding in the CrI$_2$ crystal \cite{dftd3}. We used the orthorhombic crystal structure (space group \# 36, $Cmc2_1$) for our DFT$+U$ calculation. In this space group, there are two unique crystallographic sites for the I atoms, whereas there is only one unique site for the Cr atoms. All atoms occupy the Wyckoff site labelled ``4a''. 
We performed full crystal structure relaxation (with and without the vdW dispersion correction) where both the unit cell parameters and atomic coordinates are relaxed.
Convergence thresholds for the maximum residual force and maximum energy difference between consecutive iterations were set at less than $1.5\times10^{-5}$ a.u. and $7 \times 10^{-10}$~Ry, respectively.
Representative unit cells of DFT$+U$ relaxed CrI$_2$ crystal structures are given in Fig.~\ref{fig:select_crystal}. 
The atomic parameters of the DFT$+U$ relaxed crystal structures with and without the vdW dispersion correction are in the CIF files included in the Supplemental Information.
The DFT$+U$ calculated absolute Cr magnetic moment 
is determined to be 3.79$\mu_B$.

We employed the methodology introduced by Banks et al.\ \cite{banksMagneticOrderingFrustrated2009} to compute the spin-exchange parameters, $J_n$. We constructed a 4$\times$1$\times$1 supercell and considered six independent collinear spin arrangements [one ferromagnetic (FM) and five antiferromagnetic (AFM), which are shown in Fig.\ \ref{fig:J_configures}]. We then performed self-consistent field calculations using a $\Gamma$-centered $k$-mesh of $4\times7\times4$. The total energy data, along with the Heisenberg Hamiltonian, used to calculate each $J$ from the supercell is given in subsection~\ref{sec:calculateJ} in the Supplemental Material. 

\section{Results}

\subsection{Single-crystal X-ray diffraction}

\begin{table}[t]
\caption{Lattice parameters from SCXRD refinement.}
\label{tab:SCXRD}
\begin{ruledtabular}
\begin{tabular}{lllll}
 & 100 K & 200 K & 300 K & 400 K  \\
\hline
$a$ (\AA) & 3.9070(5) & 3.9158(3) & 3.9195(3) & 3.9214(4)\\
$b$ (\AA) & 7.4963(17) & 7.5366(8) & 7.5722(6) & 7.6110(10)\\
$c$ (\AA) & 13.479(3) & 13.5142(13) & 13.5618(10) & 13.6092(17)\\
\end{tabular}
\end{ruledtabular}
\end{table}

\begin{table}[t]
\caption{Atomic positions from SCXRD refinement. Lattice parameters are shown in Table \ref{tab:SCXRD}. The space group is $Cmc2_1$. All Wyckoff symbols are ``4a''.}
\label{tab:SCXRD2}
\begin{ruledtabular}
\begin{tabular}{lllll}
\hline
& x & y & z & $U_{\textrm{iso}}$\\
\hline
100 K & & & &  \\
I1 & 0 & 0.2045(3) & 0.37815(17) & 0.0108(8) \\
I2 & 0 & 0.4720(3) & 0.62193(11) & 0.0115(8)\\
Cr1 & 0.5 & 0.3377(9) & 0.4987(6) & 0.0128(14) \\
\hline
200 K & & & & \\
I1 & 0 & 0.20577(8) & 0.37850(4) & 0.01277(17) \\
I2 & 0 & 0.47094(8) & 0.62147(3) & 0.01374(18) \\
Cr1 & 0.5 & 0.3379(3) & 0.49992(15) & 0.0156(3) \\
\hline
300 K & & & & \\
I1 & 0 & 0.20706(10) & 0.37872(5) & 0.01927(19) \\
I2 & 0 & 0.47003(9) & 0.62127(3) & 0.02039(19) \\
Cr1 & 0.5 & 0.3383(3) & 0.49975(17) & 0.0227(4) \\
\hline
400 K & & & & \\
I1 & 0 & 0.20813(16) & 0.37891(8) & 0.0270(3) \\
I2 & 0 & 0.46917(16) & 0.62096(5) & 0.0279(3) \\
Cr1 & 0.5 & 0.3378(5) & 0.5000(3) & 0.0309(7) \\
\end{tabular}
\end{ruledtabular}
\end{table}

\begin{figure}[t]
\begin{center}
\includegraphics[width=8.6cm]
{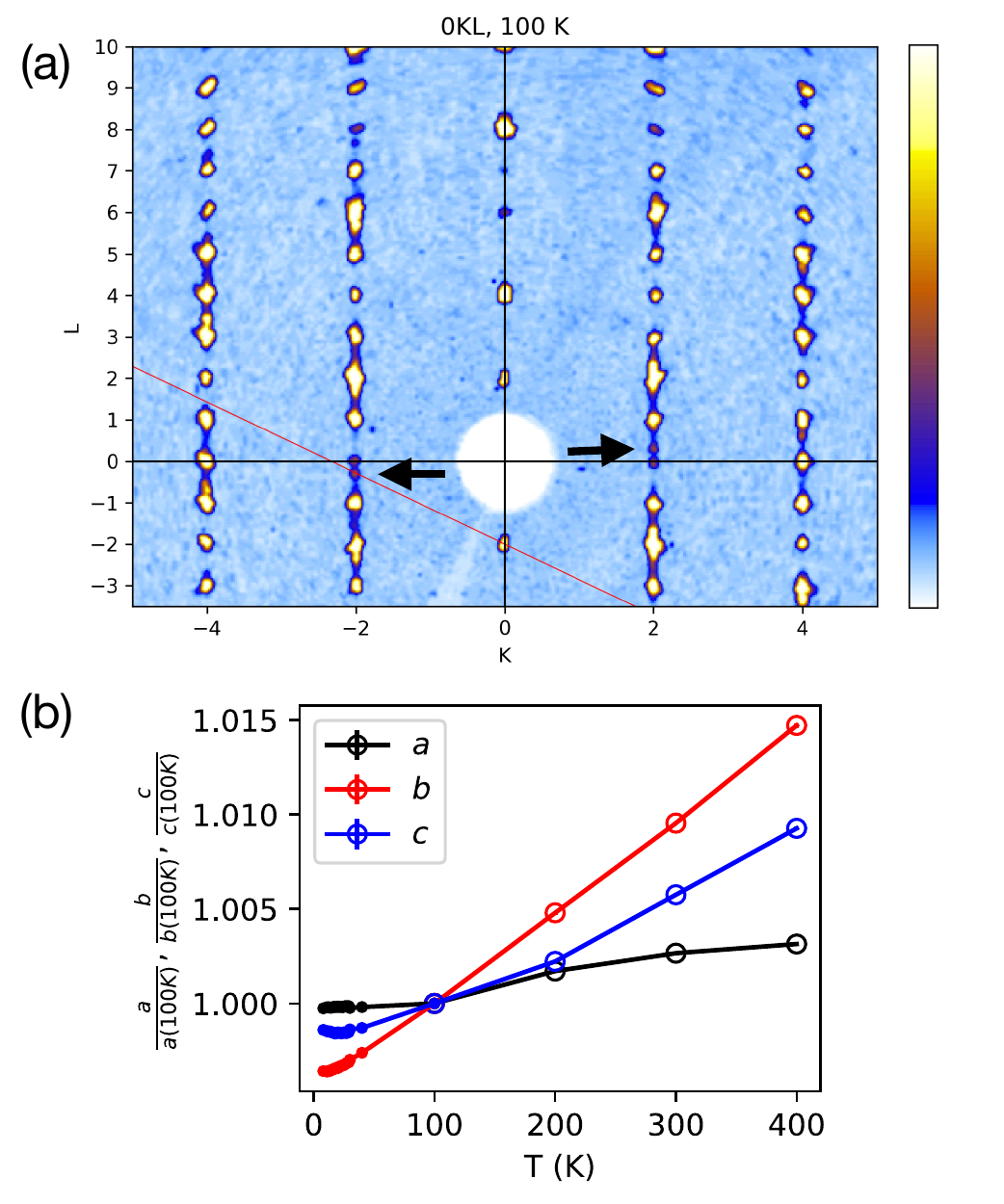}
\end{center}
\caption{(a) Single-crystal x-ray diffraction (SCXRD) intensity in the 0KL plane. In addition to the orthorhombic phase peaks, additional peaks from the monoclinic polytype \cite{tracyCrystalStructureChromium1962} are seen, with two indicated by arrows. The thin red line shows the monoclinic-phase $(0K\bar{1})$ line. (b) Lattice parameters plotted vs.\ temperature, normalized to their 100 K values. Open circles denote SCXRD values, measured for $T \geq 100$ K, and dots denote neutron diffraction values, measured for $T \leq 100$ K.}
\label{fig:SuppXRD}
\end{figure}

Single-crystal X-ray diffraction (SCXRD) on a vapor-tranport grown crystal was performed as a function of temperature. An intensity plot in the $(0KL)$ scattering plane at 100 K is shown in Fig.\ \ref{fig:SuppXRD}(a).
The refined lattice and atomic parameters are shown in Tables \ref{tab:SCXRD} and \ref{tab:SCXRD2}, respectively. From Fig.\ \ref{fig:SuppXRD}(a), not only is the main (orthorhombic) phase observed, but also a secondary, co-aligned phase with the monoclinic structure \cite{tracyCrystalStructureChromium1962} (with two of its Bragg peaks indicated by arrows), as well as diffuse scattering streaks indicating some degree of stacking disorder. No sign of a structural phase transition was observed from 100 to 400 K. A crystal from a melt-grown batch was also measured by SCXRD, and verified to have the orthorhombic structure. 
The temperature dependence of the lattice parameters (from the SCXRD and neutron diffraction data) are plotted in Fig.\ \ref{fig:SuppXRD}(b). The thermal expansion is smallest along the $a$-axis, which is as expected given the stronger bonding along the chains (as seen in structurally similar CuCl$_2$ \cite{banksMagneticOrderingFrustrated2009}.)

\subsection{Powder neutron diffraction}

\begin{figure*}[t]
\begin{center}
\includegraphics[width=17cm]
{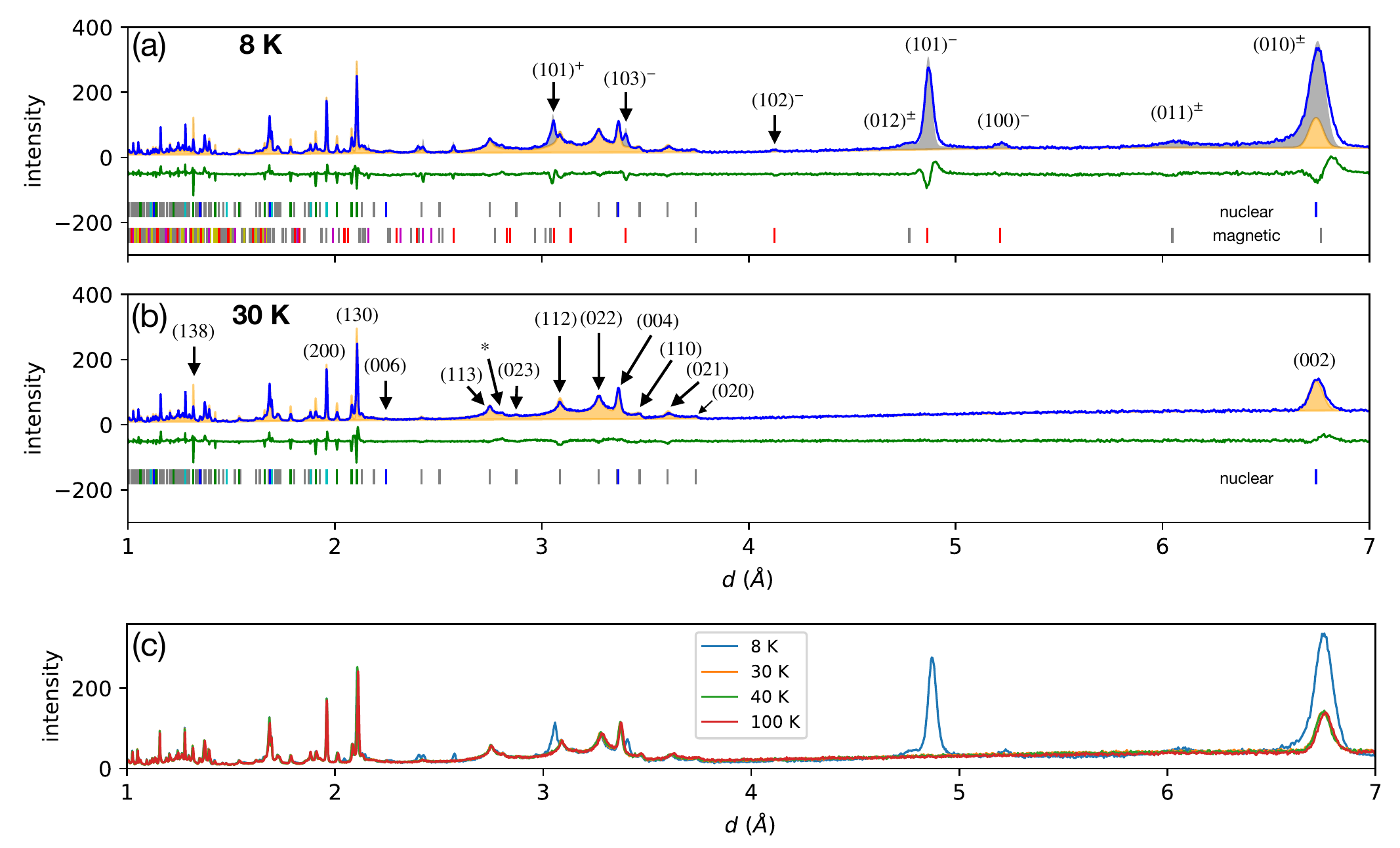}
\end{center}
\caption{(a,b) Neutron diffraction intensity (blue curves) at (a) 8 K and (b) 30 K, compared to calculated intensity from the model described in the text, shaded orange and gray for nuclear and magnetic intensity, respectively. The green curve is the difference between the data and the calculation. The locations of nuclear and magnetic peaks (with $+ / - / \pm$ superscripts omitted) are indicated below the data, colored blue for $(00L)$, magenta for $(03L)$, red for $(10L)$, green for $(13L)$, cyan for $(20L)$, yellow for $(23L)$, and gray for the remaining peaks. Selected peaks are labeled. The asterisk (*) denotes the possible $(111)$ peak of the monoclinic polytype.
(c) Neutron diffraction data at 8, 30, 40, and 100 K.
No peaks were present above $d = 7$ \AA.}
\label{fig:diffraction}
\end{figure*}

Fig.\ \ref{fig:diffraction}(a,b) is a plot of the neutron diffraction data collected at two temperatures, 8 and 30 K. A substantial amount of diffuse scattering arising from stacking disorder is observed, which is accounted for in the calculated curves and will be discussed below. Selected nuclear and magnetic Bragg peaks are labeled in Fig.\ \ref{fig:diffraction}(a,b); a more comprehensive indexing can be found in Fig.\ \ref{fig:SuppDiffraction} of the Supplemental Materials, along with the expected intensity in the absence of stacking disorder.  
At 30 K, almost every peak can be indexed by the orthorhombic (ortho-CrI$_2$) structure \cite{besrestStructureCristallineIodure1973}. Only one small peak above $d=2.0$ \AA\ (at $d=2.81$ \AA) is inconsistent with the ortho-CrI$_2$ structure; this peak is likely the $(111)$ peak of the monoclinic CrI$_2$ polytype \cite{tracyCrystalStructureChromium1962}, which would imply a volume ratio of about 1\% relative to the orthorhombic phase,  assuming no monoclinic-phase stacking disorder. 
Little change in intensity is seen from 30 to 40 K or 100 K (Fig.\ \ref{fig:diffraction}(c).)
No sign of CrI$_3$ was observed, in either the $R\bar{3}$ or $C2/m$ phases \cite{schneelochAntiferromagneticferromagneticHomostructuresDirac2023}. Some preferred orientation was present, as seen in the ratio of peaks such as $(200)$ and $(00L)$ ($L=2, 4, 6$), whose intensities would be immune to stacking disorder. We found a March-Dollase ratio of 1.15 to be satisfactory in our calculations \cite{dollaseCorrectionIntensitiesPreferred1986}.

The diffuse scattering arises from a disordered sequence of the two stacking options present in the orthorhombic CrI$_2$ crystal structure (Fig.\ \ref{fig:1}(c,d).) Using the same notation as for similar stacking variations seen in Mo$_{1-x}$W$_x$Te$_2$ \cite{schneelochEvolutionStructuralTransition2020}, we can construct one ortho-CrI$_2$ twin by a repeated ``A''-type stacking operation. Since each layer has approximate inversion symmetry, an inversion operation keeps the intralayer structure largely unchanged but changes the layer stacking from A-type to ``B''-type, related by alternating $b$-axis translations of about $\pm$0.344 lattice units.
Calculations of the diffuse scattering were performed using the formalism of Ref.\ \cite{treacyGeneralRecursionMethod1991}, which is the basis for the software \textsc{DIFFaX}; see the Supplementary Materials for mathematical details. 
For our data, we find satisfactory agreement if we assume the intensity is the sum of two contributions with 22\% of (nearly-ordered) $p=0.05$ and 78\% of (random) $p=0.5$, where $p$ is the probability of switching stacking types (A- or B-type) between consecutive layers.   
Certain peaks are expected to be unaffected by stacking disorder, in particular, $(HKL)$ where $K$ is a small multiple of 3. This invariance is due to the stacking translations randomly contributing phases of $\exp(\pm 2 \pi i K \times 0.344)$ to terms in the summation of the structure factor, which are about unity when $K$ is a small multiple of 3. The $(H0L)$ peaks are certainly prominent in our data. The $(13L)$ peaks, though, while prominent, are substantially weaker than expected from our calculations (e.g., (138) in Fig.\ \ref{fig:diffraction}(b).) A likely possibility is the presence of additional types of stacking, such as those in monoclinic CrI$_2$ \cite{tracyCrystalStructureChromium1962} or CrBr$_2$ \cite{tracyCrystalStructureChromium1962a}; preliminary calculations suggest that such stacking may be (randomly) present on the order of 10-20\%, but an in-depth calculation is beyond the scope of this paper.

\begin{figure}[t]
\begin{center}
\includegraphics[width=8.6cm]
{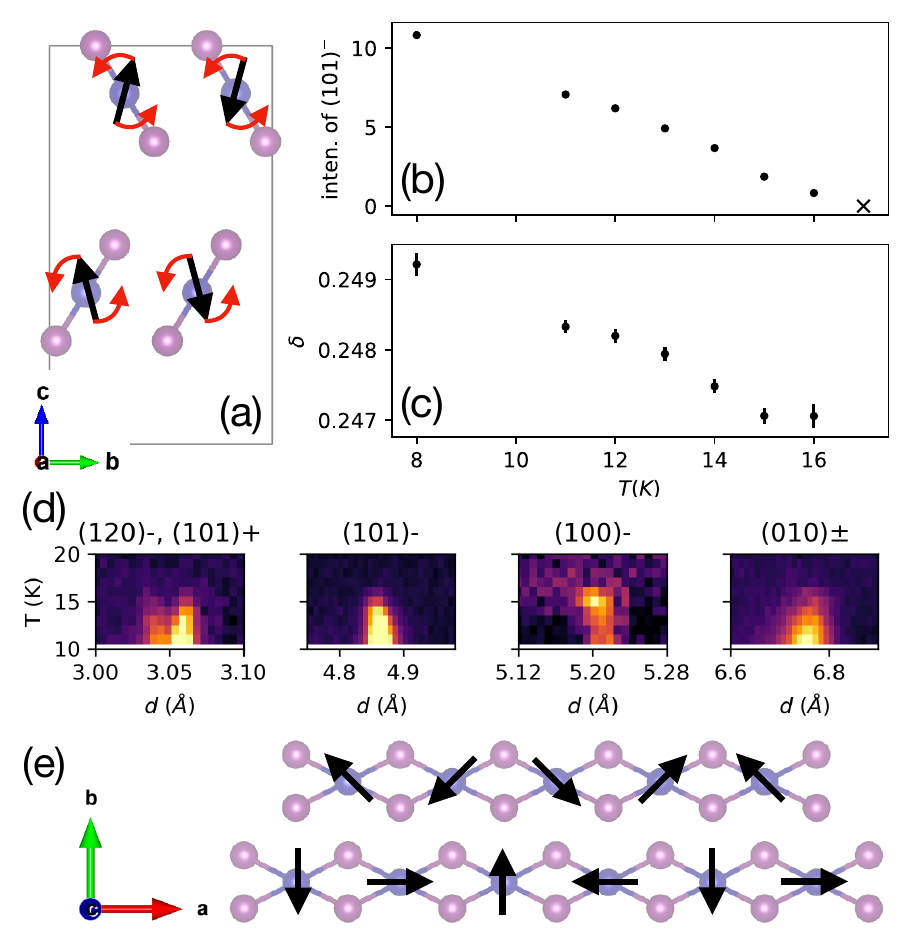}
\end{center}
\caption{(a) Depiction of the proposed ``untwisted'' spin structure in the $bc$ plane, i.e., the orientation of the spins before applying the helimagnetic spin rotation along the chain direction. (b,c) The temperature dependence of (b) the intensity of the magnetic $(101)^{-}$ peak (divided by the intensity of $(200)$) and (c) the magnitude of the modulation vector $(\delta,0,0)$. The ``X'' in (b) denotes zero intensity. (d) Magnetic intensity near selected peaks, obtained from the 11 to 20 K data by subtracting a background of the average intensity from 20 to 29 K. (e) Depiction of helical spin rotation along $a$-axis for a pair of neighboring ribbon chains within a layer. For visual clarity, these spins are depicted rotating normal to the layers, but the helical plane determined from our data is $bc$.}
\label{fig:magTDep}
\end{figure}

Below 17 K, magnetic satellite peaks appear at $(H\pm \delta, K, L)$. The peaks are labeled $(HKL)^{+}$, $(HKL)^{-}$, or $(0KL)^{\pm}$.
Their position is consistent with a spin spiral propagating along the ribbon chains, with the ``untwisted'' spin orientation (i.e., without the helimagnetic $\sim$90$^{\circ}$ rotation between neighboring Cr$^{2+}$ ions) depicted in Fig.\ \ref{fig:magTDep}(a). 
The magnetic peaks are affected by stacking disorder in the same way as  the nuclear peaks, i.e., peaks with $K=0$ remain prominent, peaks with $K=3$ are diminished relative to calculations but still prominent, and the remaining peaks are typically greatly broadened. The similar broadening for nuclear and magnetic peaks suggests that only the \emph{locations} of the spins are disordered (due to random stacking translations), while disorder in the \emph{orientation} of the spins is not needed to explain our data. 

While magnetic ordering increases on cooling, the ordered magnetic moment is not close to saturation down to 8 K, as seen from the intensity of the magnetic $(101)^{-}$ peak in Fig.\ \ref{fig:magTDep}(b). A refinement (see Supplemental Materials for details) resulted in an ordered magnetic moment of 2.95(10) $\mu_B$/Cr$^{2+}$ ion. This value is well short of the 4 $\mu_B$/Cr$^{2+}$ expected for the high-spin Cr$^{2+}$ values of $g=2$ and $S=2$. 
The magnitude of the modulation vector $(\delta,0,0)$ is plotted in Fig.\ \ref{fig:magTDep}(c); $\delta$ was obtained by fitting the positions of the $(120)^{-}$, $(101)^{-}$, $(101)^{+}$, $(100)^{-}$, $(103)^{+}$, and $(103)^{-}$ peaks, all but $(120)^{-}$ which should be unaffected by stacking disorder. 
Though $\delta \approx 1/4$, the modulation vector is incommensurate, with $\delta$ decreasing from $\sim$0.2492 to $\sim$0.2470 on warming from 8 to 16 K. These values correspond to rotations of spins between consecutive Cr$^{2+}$ ions along the ribbon chains of 89.7$^{\circ}$ and 88.9$^{\circ}$. 

The magnetic intensity agrees best with the data if we assume that the ``untwisted'' spin structure (i.e., without the helical twist, as in Fig.\ \ref{fig:magTDep}(a)) is AFM within the plane, and nearly FM but with an alternating tilt of  $\pm | \alpha | \approx \pm15(2)^{\circ}$ between the layers. 
The absence of the (002)$^{\pm}$ magnetic peak (which would be near $d=6.2$ \AA) is consistent with the untwisted spins within each layer being coaligned, since their contributions would then cancel out (see Eq.\ \ref{eq:magEq} in the Supplemental Materials.) 
At the same time, there is a small $(100)^{-}$ peak near $d=5.20$ \AA, which would be absent if the untwisted spins were collinear between the layers. 
The rotation plane of the spin helix is likely $bc$ (i.e., screw-like rather than cycloidal \cite{tokuraMultiferroicsSpinOrigin2014}), since that matches our data much better than $ab$ or $ac$, the other plausible alternatives given the symmetry of orthorhombic CrI$_2$; see Supplemental Materials for a comparison. 
With unpolarized neutrons, we are unable to tell the handedness of the helical rotation; presumably, domains with both chiralities are present.
Our spin structure bears a resemblance to that predicted by the authors of Ref.\ \cite{zhangStructuralPhaseTransitions2022}, although the helical modulation was not in their proposed structure.

What effect does the stacking disorder have on the magnetic order? On the one hand, we expect significant interlayer coupling, as was seen for structurally similar CrI$_3$ (with a value of $-0.59$ meV for the summed exchange constants for interlayer bonding for the $R\bar{3}$ phase \cite{chenMagneticFieldEffect2021}). Our DFT calculations (to be discussed below) also suggest a sizable interlayer magnetic coupling.
However, for symmetry reasons, we expect the magnetic ordering to be largely unaffected by the A/B stacking disorder. 
If we apply an inversion operation to one twin of the structure, which we call ``AA'' to denote its A-type stacking, we end up with the ``BB'' twin. 
However, spin is a pseudovector, and does not change its orientation upon inversion, so the same spin structure that exists for the AA twin should be valid for the BB twin.  
Thus, assuming that magnetic interactions beyond nearest neighboring layers are negligible, the relative spin orientation of neighboring layers should be independent of A- or B-type stacking, and no disorder in the orientation of the magnetic moments should arise from the A/B stacking disorder. 
This situation is different from that in CrI$_3$, where the different types of stacking (monoclinic and rhombohedral) are not symmetry-equivalent to each other, and a stacking dependence of the spin orientation is observed \cite{schneelochAntiferromagneticferromagneticHomostructuresDirac2023}.  
We note, though, that partial layer sliding, such as during coherent phonon oscillations produced in ultrafast pump-probe spectroscopy  \cite{weberUltrafastInvestigationControl2021}, would modulate the interlayer magnetic coupling, and thus the relative phase of the spin helices of neighboring layers.

\begin{figure}[t]
\begin{center}
\includegraphics[width=8.6cm]
{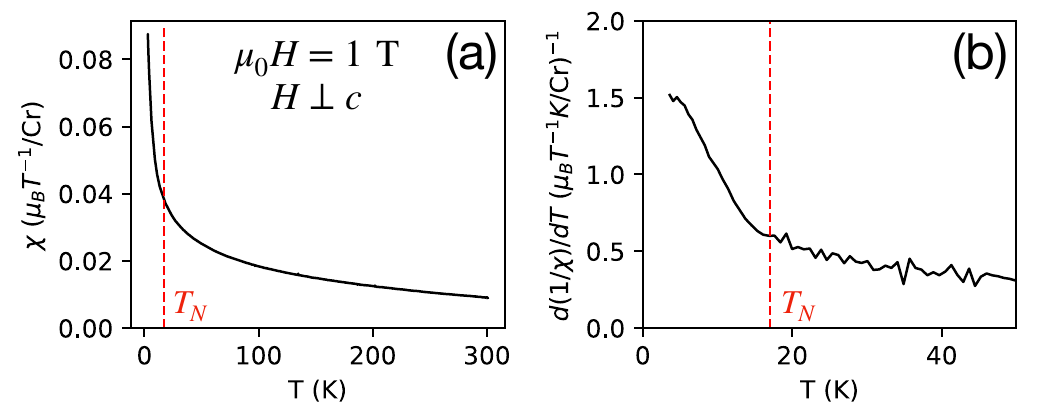}
\end{center}
\caption{(a) Magnetic susceptibility of a melt-grown CrI$_2$ crystal with a 1 T magnetic field applied in-plane. (b) The derivative $d(1/\chi)/dT$ for the data in (a). 
$T_N = 17$ K is depicted as a red dashed line.}
\label{fig:magnetization}
\end{figure}

\subsection{Magnetic susceptibility}

The onset of helimagnetic order can also be inferred from magnetization data taken on melt-grown crystals, suggesting that the magnetic transition is, at most, weakly affected by stacking disorder. 
The magnetization has a monotonic increase on cooling (Fig.\ \ref{fig:magnetization}(a).) 
This behavior contrasts with the broad hump often seen in quasi-1-dimensional magnetic systems, with the most relevant examples being CrCl$_2$ \cite{hagiwaraMagneticPropertiesAnhydrous1995}, CuCl$_2$ \cite{banksMagneticOrderingFrustrated2009}, and CuBr$_2$ \cite{zhaoCuBr2NewMultiferroic2012,leeInvestigationSpinExchange2012}, suggesting the role of stronger inter-layer and inter-chain interactions in CrI$_2$. 
The inverse magnetization is roughly linear above $T_N\sim17$ K, with a fit resulting in a Curie-Weiss temperature of about -70 K (see Fig.\ \ref{fig:magSupp1}(a) in the Supplemental Materials), suggesting the dominance of antiferromagnetic interactions. Below $T_N$, though, the derivative $d(1/\chi)/dT$ shows a marked upturn, indicating the onset of helimagnetic ordering (Fig.\ \ref{fig:magnetization}(b).) 
Magnetization vs.\ field at 3 K (Fig.\ \ref{fig:magSupp1}(b)) shows linear and non-hysteretic behavior at low fields, with a possible transition induced by fields above $\sim$3 T. Generally, we saw similar behavior for magnetic field applied in-plane or out-of-plane, but further study is needed to determine the extent to which in-plane anisotropy is present, i.e., along or perpendicular to the ribbon chains.

% DFT CALCULATIONS
\begin{figure}[t]
\begin{center}
\includegraphics[width=8.6cm]
{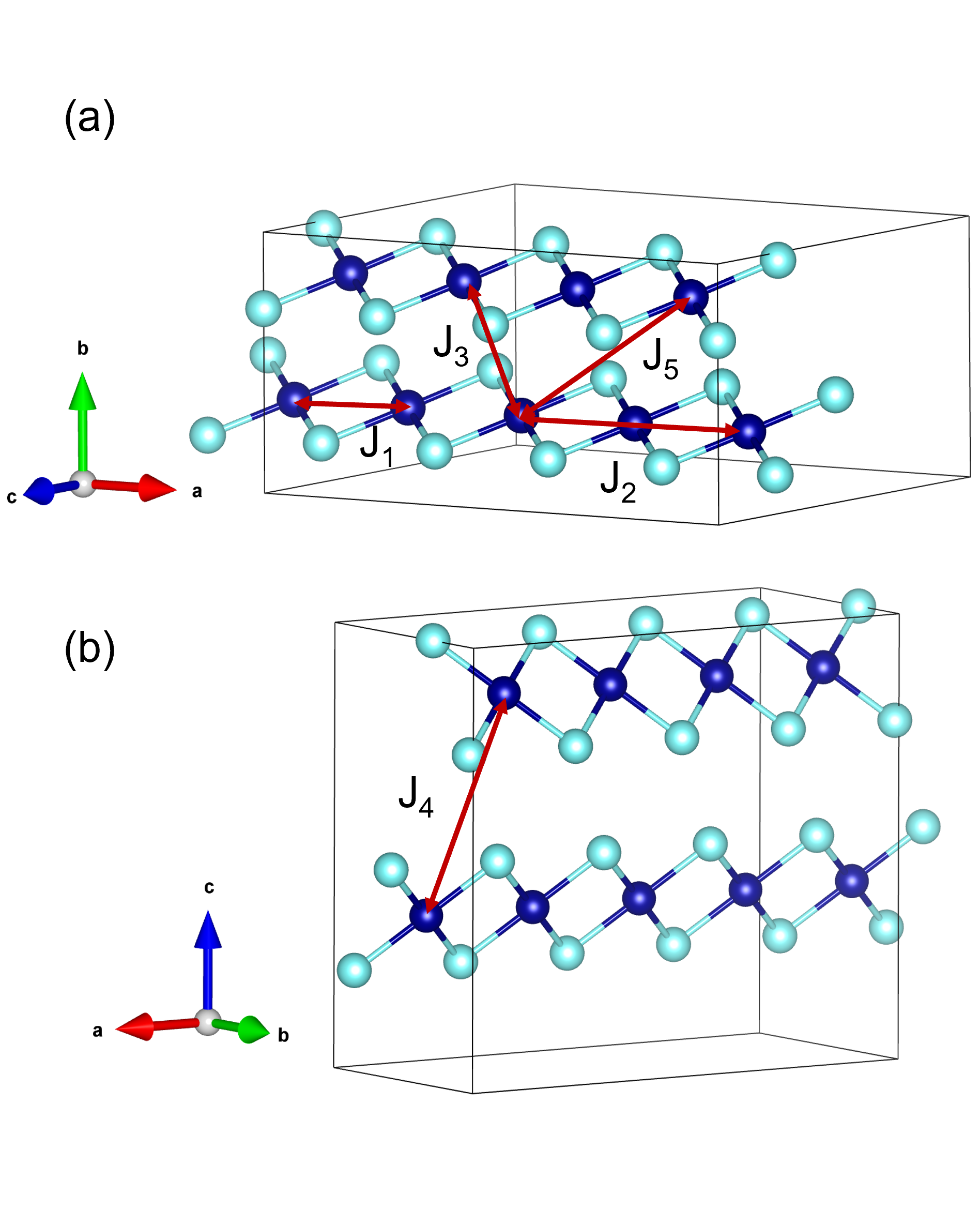}
\end{center}
\caption{Definition of $J_n$ of CrI$_2$ supercell. The dark blue color represents Cr atoms, and the cyan color corresponds to I atoms. (a) $J_1$ and $J_2$ is nearest- and next-nearest-neighbor intrachain Cr-Cr interaction. $J_3$ and $J_5$ is nearest- and next-nearest-neighbor intralayer interchain Cr-Cr interaction. (b) $J_4$ is nearest-neighbor interlayer Cr-Cr interaction.}
\label{fig:forDFT}
\end{figure}

% values per Cr-Cr bond:
\begin{table}[t]
\caption{Exchange interactions per Cr-Cr bond from DFT$+U$ calculations ($U=3$~eV) using crystal structures that were relaxed with and without vdW correction. $J_n<0$ indicates favoring AFM exchange coupling.}
\label{tab:DFT}
\begin{ruledtabular}
\begin{tabular}{lll}
Relaxation procedure & Cr-Cr (\AA) & $J_n$ (meV) \\
\hline
& 3.930 & $J_1$=2.99 \\
 & 7.861 & $J_2$=$-$2.19 \\
 DFT+$U$+vdW relaxed & 4.196 & $J_3$=$-$0.42 \\
 & 7.017 & $J_4$=2.11 \\
 & 6.965 & $J_5$=0.03 \\
 \hline \hline
 & 3.947 & $J_1$=3.55 \\
 & 7.894 & $J_2$=$-$2.14 \\
DFT+$U$ relaxed & 4.279 & $J_3$=$-$0.27 \\
 & 7.195 & $J_4$=1.36 \\
 & 7.033 & $J_5$=0.03 \\
\end{tabular}
\end{ruledtabular}
\end{table}

To determine whether or not the helimagnetism of CrI$_2$ has a similar origin as for the intrachain frustrated magnetic interactions in CuX$_2$ (X=Cl, Br),  we  performed DFT$+U$ calculations using crystal structures that were relaxed with and without the vdW dispersion correction. We computed the energies of a number of collinear spin configurations, then mapped their relative energies to the following spin Hamiltonian,
\begin{equation}
    \label{eq:spinHamiltonian}
H = \sum_{i<j} -J_{ij} \mathbf{\hat{S}}_i \cdot \mathbf{\hat{S}}_j
\end{equation}
to obtain the exchange interactions $J_{n}$ (our definition of $J_1$--$J_5$ is shown in Fig.\ \ref{fig:forDFT}.) 
In terms of differences in the DFT$+U$ relaxed crystal structures, introduction of vdW dispersion correction has an impact on the unit cell volume. The crystal structure with vdW correction had a smaller unit cell volume (385~{\AA}$^3$) when compared to the crystal structure without the vdW dispersion correction (404~{\AA}$^3$). As a result, introduction of vdW correction reduced the Cr-Cr bond lengths leading to closer intrachain and interlayers bond distances (see Table~\ref{tab:DFT}). 

The DFT$+U$ derived exchange interactions for coordinates relaxed with or without the vdW correction are shown in Table~\ref{tab:DFT}. The total energy difference data used for the calculation of $J_n$ is given in Table~\ref{tab:J_cal}. 
As shown in Fig.~\ref{fig:forDFT}, the terms $J_1$ and $J_2$ correspond to exchange energies that arise due to the nearest- and next-nearest-neighbor intrachain Cr-Cr bonds. 
According to the $J_1$-$J_2$ model \cite{blundellMagnetismCondensedMatter2001}, the necessary condition for helimagnetism is $|J_2| > |J_1|/4$. 
From Table~\ref{tab:DFT}, our DFT$+U$ calculated $J_1$ and $J_2$ satisfies the $J_1$-$J_2$ model condition (irrespective of the choice of crystal structure), which supports the formation of helimagnetism in CrI$_2$ crystals. 
We attribute this to the sizable AFM next-nearest-neighbor $J_2$ intrachain interaction term.
We also infer from the magnitude of $|J_4|$ that the inter-layer magnetic coupling is non-negligible and cannot be ignored. 
For the intralayer-interchain interactions, the nearest- and next-nearest-neighbor Cr-Cr bonds correspond to $J_3$ and $J_5$, respectively, and they are smaller than that of the $J_1$, $J_2$ and $J_4$ parameters.

\section{Discussion}

There are striking similarities in both the structure and magnetism between CrI$_2$ and CuX$_2$ (X=Cl, Br), as well as many other Cu$^{2+}$ ribbon-chain compounds. Aside from differences in layer stacking, CuX$_2$ (X=Cl, Br) is isostructural to CrX$_2$ (X=Br, I), with JT distortions of the Cu$^{2+}$X$^{-}_6$ and Cr$^{2+}$X$^{-}_6$ octahedra resulting in ribbon-chain structures. Helimagnetic order has been reported in CuCl$_2$ below 24 K \cite{banksMagneticOrderingFrustrated2009}, and in CuBr$_2$ below 73.5 K \cite{zhaoCuBr2NewMultiferroic2012}, with both exhibiting $\sim$90$^{\circ}$ rotations of spins along the chain axis between consecutive Cu/Cr ions (specifically, 81$^{\circ}$ for CuCl$_2$ and 85$^{\circ}$ for CuBr$_2$), and an antiferromagnetism between intralayer chains similar to CrI$_2$. 
One difference, though, is the plane of the helical rotation; CuCl$_2$ and CuBr$_2$ are reported to have spins rotate cycloidally within the plane that includes the chain and out-of-plane directions, whereas the spins in orthorhombic CrI$_2$ appear to rotate perpendicular to the chain. A number of other Cu$^{2+}$ ribbon-chain compounds also exhibit helimagnetism, such as LiCuVO$_4$ \cite{enderleQuantumHelimagnetismFrustrated2005,gibsonIncommensurateAntiferromagneticOrder2004}, LiCu$_2$O$_2$ \cite{masudaCompetitionHelimagnetismCommensurate2004}, NaCuMoO$_4$(OH) \cite{asaiHelicalCollinearSpin2020}, and PbCuSO$_4$(OH)$_2$ \cite{willenbergMagneticFrustrationQuantum2012}. 
The helimagnetism is often described in terms of the $J_1$-$J_2$ model. 
It is frequently assumed that the helimagnetism in the Cu$^{2+}$ ribbon chain compounds arises from the frustration between an AFM $J_2$ and a sufficiently weak $J_1$, though there are disagreements as to whether the $J_1$-$J_2$ model can fully explain the magnetism or where different materials fall in regards to the size of the $|J_1 / J_2|$ ratio. 
While there has been plenty of research into the helimagnetism of Cu$^{2+}$ compounds, few such studies have been done of their Cr$^{2+}$ counterparts. 
(CrCl$_2$ is an exception; though collinear AFM ordering sets in below 16 K \cite{cableNeutronDiffractionStudies1960} or 11.3 K \cite{winkelmannStructuralMagneticCharacterization1997}, depending on the polymorph, intriguingly, short-range helimagnetic order has been reported \cite{winkelmannStructuralMagneticCharacterization1997}, with a $\sim$97$^{\circ}$ rotation angle between neighboring spins.)
The case of CrI$_2$ suggests a similarity in the magnetism of Cr$^{2+}$ and Cu$^{2+}$ ribbon chain compounds that should be explored further.

A better understanding of the connection between the helimagnetism and ribbon-chain structure of CrI$_2$ could be achieved by 1) measuring the magnetic behavior of CrBr$_2$, which is, thus far, unreported, 2) performing a more detailed study on the short-range helimagnetism of CrCl$_2$, and 3) conducting inelastic neutron scattering measurements on CrI$_2$ to validate the exchange interactions between the ions, paired with more in-depth DFT$+U$ calculations that go beyond collinear spin structures and include the effects of spin-orbit coupling. 

The helimagnetism of CrI$_2$ presents a number of opportunities for future research. 
There is the potential for sliding ferroelectricity in orthorhombic CrI$_2$ by virtue of its layer stacking \cite{zhangStructuralPhaseTransitions2022}, in which the electric polarization may be switchable via layer sliding as has been observed for other bilayer systems such as WTe$_2$ and hexagonal boron nitride \cite{wuSlidingFerroelectricity2D2021}. 
Another research opportunity involves the search for a monolayer helimagnetic multiferroic. NiI$_2$ has recently gained attention for being a potential single-layer multiferroic \cite{songEvidenceSinglelayerVan2022}, where the ferroelectricity (FE) is assumed to arise from the helimagnetic ordering \cite{tokuraMultiferroicsSpinOrigin2014} (though there is controversy as to whether FE has, in fact, been demonstrated \cite{jiangDilemmaOpticalIdentification2023}.) We have shown that CrI$_2$ is helimagnetic in the bulk, and isolation of monolayers of CrI$_2$ has already been reported \cite{caiMolecularBeamEpitaxy2021,liTwoDimensionalMagneticSemiconducting2023,pengMottPhaseVan2020}, so monolayer CrI$_2$ would be a natural candidate for helimagnetism. 
We should make a couple of points, however. 
First, while (nearly-isostructural) CuBr$_2$ was reported to have strong magnetoelectric coupling associated with its helimagnetic ordering \cite{zhaoCuBr2NewMultiferroic2012}, the spin helix is cycloidal in CuBr$_2$ and CrI$_2$, but screw-like in CrI$_2$, and only cycloidal helices are expected to generate a ferroelectric polarization by the inverse Dzyaloshinskii-Moriya mechanism \cite{tokuraMultiferroicsSpinOrigin2014}. However, an applied magnetic field may change the helical plane and allow (or suppress) FE, as has been demonstrated for CuCl$_2$ \cite{sekiCupricChlorideTextCuCl2010}. 
Second, the magnetic order for monolayer CrI$_2$ may differ from that of the bulk; the helimagnetism in monolayer NiI$_2$, in fact, has been reported to have a different wavevector than bulk NiI$_2$ \cite{miaoSpinresolvedImagingAtomicscale2023}.
An investigation into the magnetic properties of monolayer CrI$_2$ would, no doubt, enlighten the search for novel magnetic materials in the few-layer limit.

\section{Conclusion}

In conclusion, we observed via neutron diffraction the onset of helimagnetic order below 17 K in orthorhombic CrI$_2$. The spin helix propagates along the chain direction with a wavevector of about $(0.2492,0,0)$ at 8 K, decreasing slightly in magnitude on warming. Our DFT$+U$ calculations provide evidence suggesting that one of the key factors behind the helimagnetism may be a sizable antiferromagnetic next-nearest-neighbor intrachain coupling.

\section*{Acknowledgements}

The work at the University of Virginia is supported by the Department of Energy, Grant number DE-FG02-01ER45927. A portion of this research used resources at the Spallation Neutron Source, a DOE Office of Science User Facility operated by Oak Ridge National Laboratory.
DFT$+U$ calculations are performed using the Rivanna cluster that is supported and maintained by the University of Virginia research computing.

% \bibliography{CrI2Paper}


\newpage
\section{Supplemental Materials}
\beginsupplement

\subsection{Ideal diffraction pattern and Bragg peak locations}
\begin{figure*}[t]
\begin{center}
\includegraphics[width=17cm]
{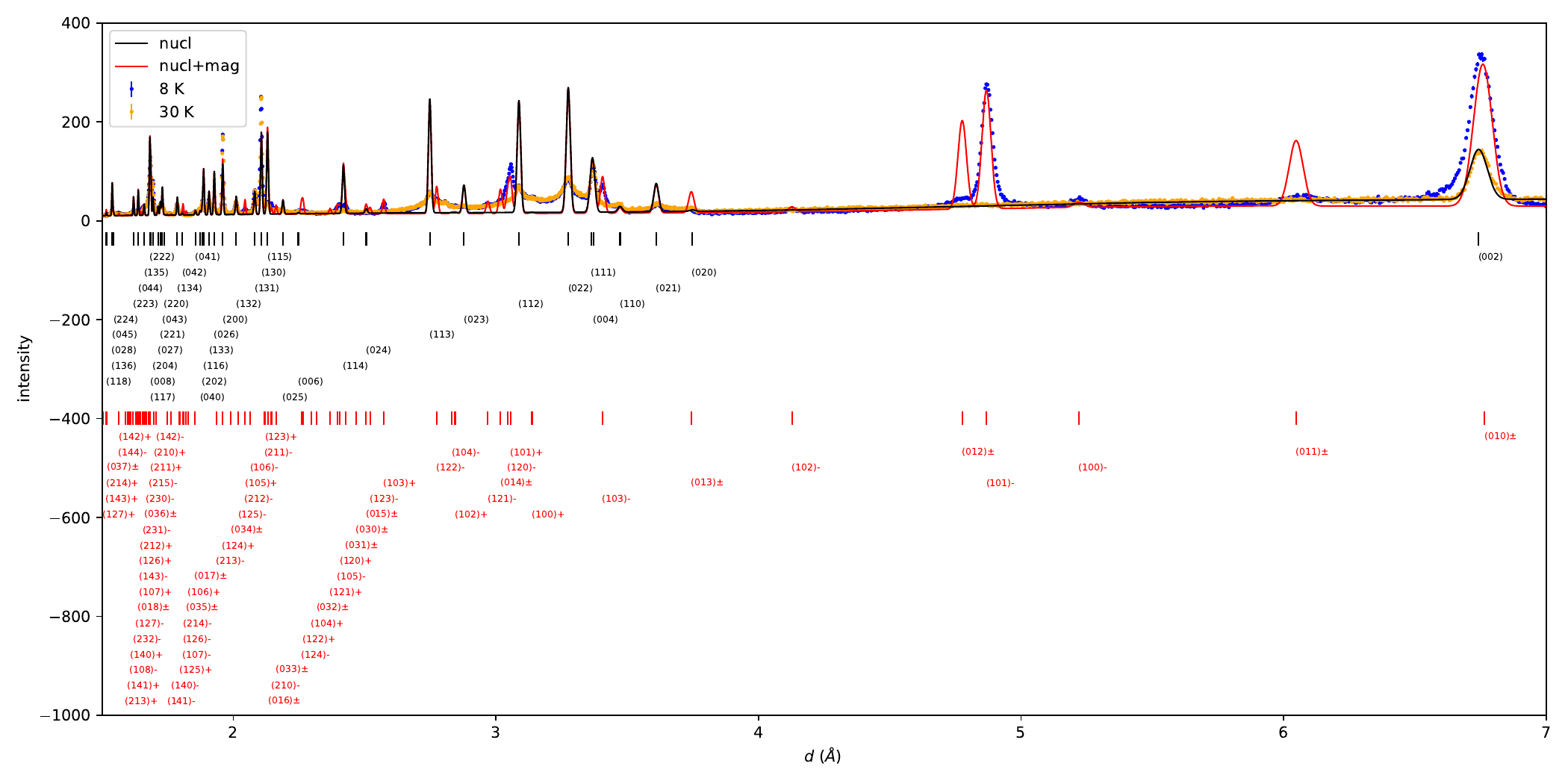}
\end{center}
\caption{Neutron diffraction data at 8 and 30 K, with calculated intensity assuming no stacking disorder for both nuclear (nucl) scattering and both nuclear and magnetic (nucl+mag) scattering. The nuclear Bragg peak locations for orthorhombic CrI$_2$ are shown in black, and the magnetic peaks are shown below in red.}
\label{fig:SuppDiffraction}
\end{figure*}

In Fig.\ \ref{fig:SuppDiffraction}, we show the 8 and 30 K data, along with calculated intensity for an ideal orthorhombic CrI$_2$ structure, both for purely nuclear neutron scattering and with magnetic scattering included. The nuclear and magnetic peak locations are depicted underneath.

\subsection{Computation of the diffuse scattering}

Here, we discuss our usage of the formalism of Ref.\ \cite{treacyGeneralRecursionMethod1991} to compute the diffuse scattering intensity. Although \textsc{DIFFaX}, the software based on this formalism, cannot do magnetic neutron scattering intensity calculations, it is straightforward to implement the formalism for this purpose. 

For nuclear neutron scattering, the layer form factor $F_l(\mathbf{Q})$ is defined as 
\begin{equation}
    F_l(\mathbf{Q}) = \sum_j b_j exp(i \mathbf{Q} \cdot \mathbf{d}_j) \exp(-W_j).
\end{equation}
The only difference from the formula for the (nuclear) structure factor is that the sum, here, is only over atoms for a single layer. The nuclear scattering length of the $j$th atom is $b_j$, $\mathbf{d}_j$ is the $j$th atom's position, $\mathbf{Q}$ is the momentum transfer, and $\exp(-W_j)$ is the Debye-Waller factor (set to  $U_{\textrm{iso}} = 0.011$ \AA$^{2}$ for all atoms.)

In the formalism used here for orthorhombic-like stacking disorder, we consider four types of layers, depending on whether the layers are oriented along the $b$-axis, and depending on whether those layers were preceded by an A- or B-type stacking boundary (which would result in different in-plane translations relative to the layer below.) Thus, we can construct a $4 \times 1$ matrix $\mathbf{F}$ of the layer form factors for the four types of layers. Next, we construct the $4 \times 4$ matrix $\mathbf{T}$. Its elements are $\alpha_{ij} \exp(-i \mathbf{Q} \cdot \mathbf{R}_{ij})$, where $\mathbf{R}_{ij}$ is the translation from a type-$i$ bottom layer to a type-$j$ layer above it (including the out-of-plane component $c \hat{z}/2$), and $\alpha_{ij}$ is the probability of having a type-$j$ upper layer given a type-$i$ bottom layer. Next, we solve $\mathbf{\Phi} = \mathbf{F} + \mathbf{T \Phi}$ for the scattered wavefunction amplitude $4 \times 1$ matrix $\mathbf{\Phi}$, which requires inverting $(\mathbb{I}-\mathbf{T})$, where $\mathbb{I}$ is the $4 \times 4$ identity matrix. To avoid singularities, we follow the advice of Treacy et al.\ \cite{treacyGeneralRecursionMethod1991}, who recommend multiplying the elements of $\alpha_{ij}$ by $(1 - 2 \sigma^2_{ij})$, with $2 \sigma^2_{ij}$ chosen to be $10^{-3}$. This ``trick'' adds an effective uncertainty to the layer translations, and its effect is to almost imperceptibly broaden peaks while allowing the calculation to converge. Finally, the intensity can be calculated via $(\mathbf{G^{*T} \Phi + G^{T} \Phi^{*} - G^{*T} F})$, where $\mathbf{G}$ is the $4 \times 1$ matrix with elements $g_i \mathbf{F}_i(\mathbf{Q})$, with $g_i$ being the overall frequency of having layers of type $i$. $\mathbf{G}^{\mathbf{T}}$ and $\mathbf{G}^{*}$ denote the transpose and complex conjugate of $\mathbf{G}$, respectively. 
Intensity ``rods'' are  computed at positions $(H,K,L)$, with integer $H$ and $K$ and a grid of points along $L$. 
Once the intensity rods have been calculated, the intensity is powder-averaged, then convoluted with the instrument resolution function and multiplied by the Lorentz factor for time-of-flight neutron scattering to simulate the data.

For magnetic scattering, a similar method is followed. First, the magnetic structure factor of a peak, given helimagnetic ordering with a modulation vector $\mathbf{q}$, is \cite{kuindersmaMagneticStructuralInvestigations1981,carpenterElementsSlowNeutronScattering2015} 
\begin{multline}
\label{eq:magEq}
\mathbf{F}_M(\mathbf{G \pm q}) = \frac{\gamma_n r_0}{2 \mu_B} \times \\ \sum_j f(\mathbf{G} \pm \mathbf{q}) \frac{|\mathbf{m}_j^{\mathbf{q}}|}{2} (\hat{u}_j \pm i \hat{v}_j) \times \\ 
\exp(i (\mathbf{G} \cdot \mathbf{d}_j \mp \phi_j)) 
\exp(-W_j).
\end{multline}
The $j$ index runs over every Cr$^{2+}$ ion in the unit cell.
Here, $\frac{\gamma_n r_0}{2 \mu_B} = 2.696$ fm, $f(\mathbf{Q})$ is the magnetic form factor (in this case, for Cr$^{2+}$), $\mathbf{G}$ is the location of $(H,K,L)$, $\mathbf{m}_j^{\mathbf{q}}$ is the Fourier component of the $j$th magnetic moment in the unit cell ($\mathbf{m}_j^{\mathbf{q}}$ are in reference to the ``untwisted'' spins of the unit cell, before applying the helical rotation), $\mathbf{d}_j$ is the position of the $j$th ion, $\phi_j$ is a phase factor for the $j$th spin, and $W_j$ is the Debye-Waller factor. 
The unit vectors $\hat{u}_j$ and $\hat{v}_j$ are orthogonal, defining the plane of rotation for the spin helix. 

The magnetic layer form factor $\mathbf{F}_{M,l}(\mathbf{Q})$ %(Eq.\ \ref{eq:magEq}) 
is defined analogously to the case of nuclear scattering, with the same form as Eq.\ \ref{eq:magEq} except that the sum runs over spins in only one layer rather than all layers in a unit cell. An important difference from nuclear scattering is that $\mathbf{F}_{M,l}(\mathbf{Q})$ is a \emph{vector} rather than a scalar \cite{shiraneNeutronScatteringTripleAxis2002}, i.e., it will have components $F^x_{M,l}(\mathbf{Q})$, $F^y_{M,l}(\mathbf{Q})$, and $F^z_{M,l}(\mathbf{Q})$. A magnetic Bragg peak measured with unpolarized neutrons will have an intensity proportional to $\mathbf{F}_M(\mathbf{G}) \cdot \mathbf{F}^*_M(\mathbf{G}) = |F^x_M(\mathbf{G})|^2 + |F^y_M(\mathbf{G})|^2 + |F^z_M(\mathbf{G})|^2$, where $\mathbf{F}_M(\mathbf{G})$ is the magnetic structure factor at a reciprocal lattice vector $\mathbf{G}$. In other words, we can determine the intensity from the sum of three contributions corresponding to the $x$, $y$, and $z$ axes. We then follow the formalism of Treacy, et al.\ \cite{treacyGeneralRecursionMethod1991}, except that we sum \emph{three} separate contributions, corresponding to the three magnetic layer form factors $F^x_{M,l}(\mathbf{Q})$, $F^y_{M,l}(\mathbf{Q})$, and $F^z_{M,l}(\mathbf{Q})$, each contribution computed analogously as for the nuclear diffuse scattering.

\subsection{Magnetic intensity calculations for different planes of helical rotation}

\begin{figure*}[t]
\begin{center}
\includegraphics[width=17cm]
{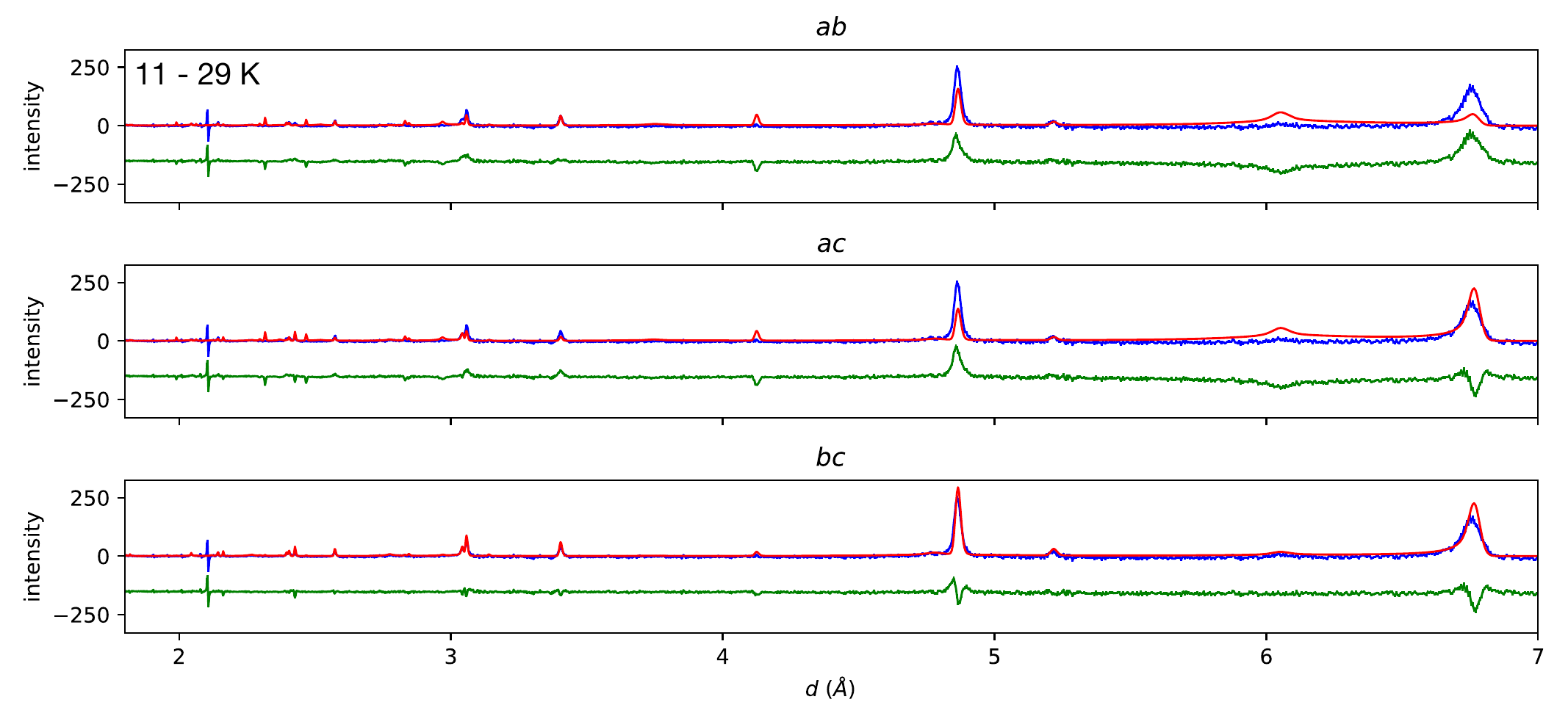}
\end{center}
\caption{Magnetic intensity (11 K minus 29 K data, blue) compared with calculated magnetic intensity (red) for helical rotation in the $ab$, $ac$, or $bc$ planes. The green curves show the difference between the data and the calculated intensity.}
\label{fig:helixPlanes}
\end{figure*}

In Fig.\ \ref{fig:helixPlanes}, we show calculated magnetic intensity using the same parameters as for Fig.\ \ref{fig:diffraction}(b) of the main text, except that only the magnetic intensity is shown (i.e., the 11 K minus 29 K intensity), and that calculations were done for three different planes of helical rotation: the $ab$, $ac$, and $bc$ planes. We see that the $bc$ plane best describes the data. In other words, the helical spin order for orthorhombic CrI$_2$ is most likely screw-like, in contrast to the cycloidal order reported for CuCl$_2$ and CuBr$_2$ \cite{banksMagneticOrderingFrustrated2009,leeInvestigationSpinExchange2012}.

\subsection{Additional magnetization data}

\begin{figure}[t]
\begin{center}
\includegraphics[width=8.6cm]
{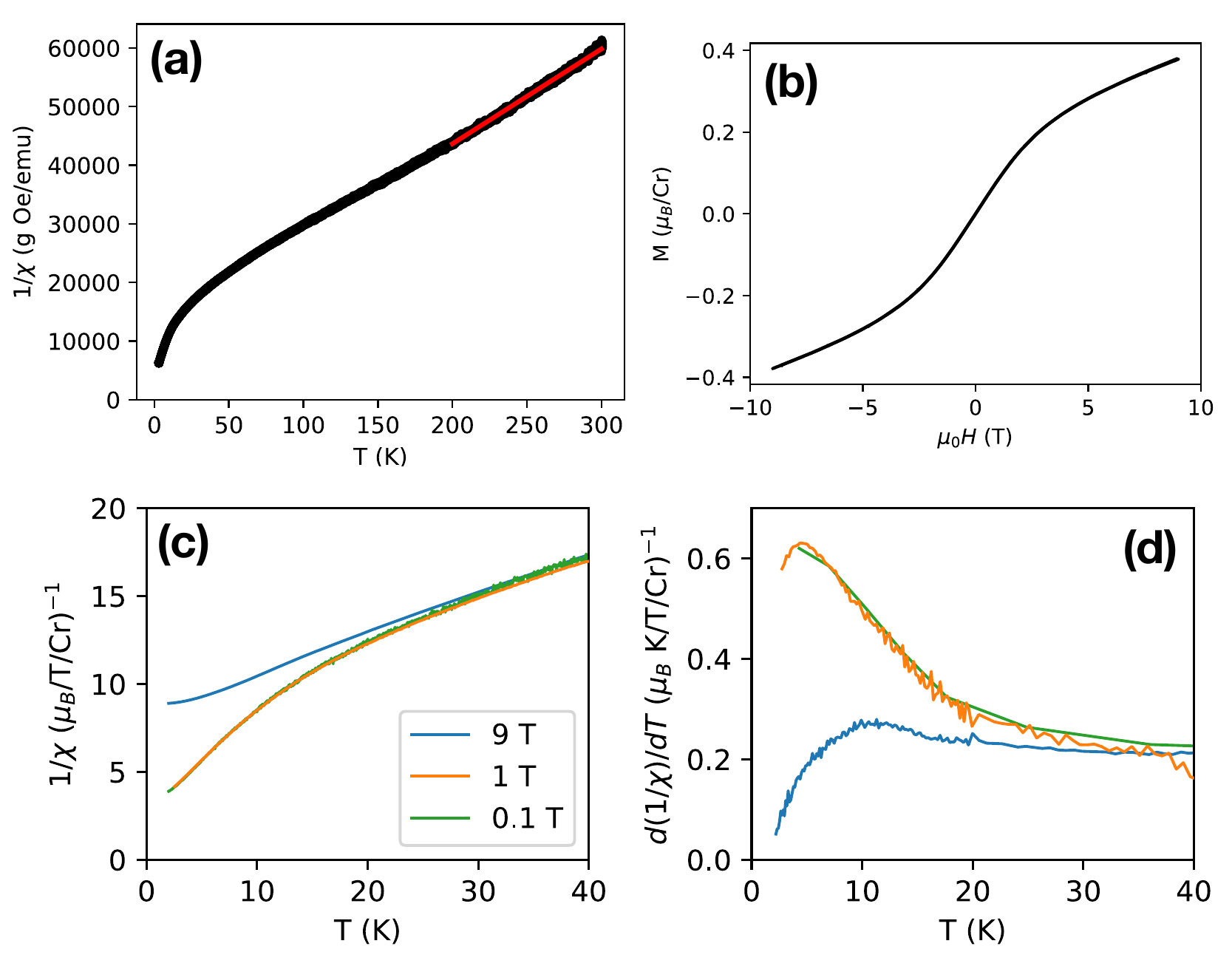}
\end{center}
\caption{(a) Inverse susceptibility for the same data shown in Fig.\ \ref{fig:magnetization}. A field of 1 T was applied. The red line shows a Curie-Weiss fit. (b) Magnetization-field hysteresis loop for the same melt-grown crystal measured for Fig.\ \ref{fig:magnetization}, taken at 3 K. (c,d) Magnetization data taken on an additional melt-grown CrI$_2$ crystal at the indicated fields. Magnetic field was applied in-plane for all subplots.}
\label{fig:magSupp1}
\end{figure}

In Fig.\ \ref{fig:magSupp1}(a), we show the inverse susceptibility for the same data as in Fig.\ \ref{fig:magnetization}. The red line indicates a Curie-Weiss fit of a line to the region $200 \leq T \leq 300$ K, from which we obtained $T_{CW} = -69.8$ K and an effective moment of $\sqrt{g^2 J(J+1)}$ = 3.89 $\mu_B$. (The fitted uncertainties were 0.4 K and 0.0025 $\mu^{\textrm{eff}}_B$, respectively, but the true uncertainties were certainly larger, perhaps due to impurities induced by water absorption; for instance, another Curie-Weiss analysis on a separate crystal (at 9 T, with in-plane field) yielded $T_{CW} =  -26.1$ K and $\mu_B^{\textrm{eff}} = $4.186 $\mu_B$.) The ideal effective moment for $g=2$, $J=S=2$ would be 4.90 $\mu_B^{\textrm{eff}}$, suggesting a deviation from Curie-Weiss behavior or the effect of impurities in our measurements. 

In Fig.\ \ref{fig:magSupp1}(b), we show a magnetization-field hysteresis loop at 3 K for the same crystal and ($\mathbf{H} \perp \mathbf{c}$) orientation as for Fig.\ \ref{fig:magnetization}. The magnetization is roughly linear in field up to 1 to 2 T, but deviates from linearity for larger fields. At 9 T, the magnetization is still well below the expected saturation value of 4 $\mu_B$/Cr for $g=2$ and $J=S=2$. In Fig.\ \ref{fig:magSupp1}(c,d) we show magnetization data for an additional melt-grown CrI$_2$ crystal, for fields at 0.1, 1, and 9 T. The inverse susceptibility in Fig.\ \ref{fig:magSupp1}(c) nearly overlaps between 0.1 and 1 T, both showing similar behavior as for Fig.\ \ref{fig:magnetization}, but the 9 T data show different behavior, suggesting that a field beyond $\sim$3 T (judging from Fig.\ \ref{fig:magSupp1}(b)) may induce a change in magnetic ordering. Neutron diffraction studies under magnetic field would clarify this behavior.

\begin{figure*}[!htb]
    \centering
    \includegraphics[width=17cm]{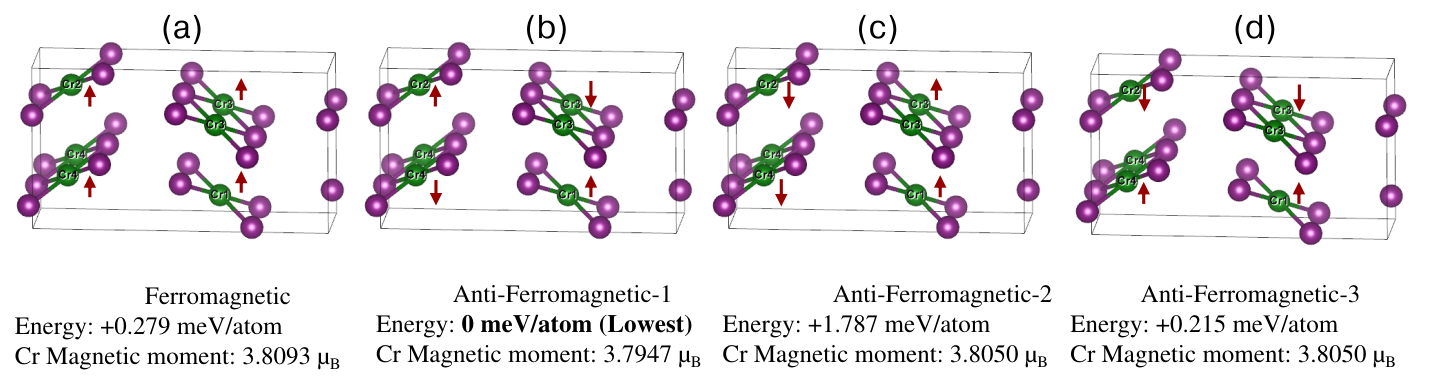}
    \caption{DFT$+U$ relaxed four different collinear spin configurations using CrI$_2$ unit cell (without vdW correction). Both the cell parameters and internal coordinates were relaxed in these calculations. In (a) we show the FM configuration. In (b), (c), and (d) three different AFM configurations are shown. Among the four configurations, the AFM configuration shown in (b) converged with the lowest total energy. The absolute value of the Cr magnetic moment for the lowest energy configuration is 3.79~$\mu_B$. The CIF file corresponding to the crystal structure shown in (b) is included in the Supplemental Material.}
    \label{fig:select_crystal}
\end{figure*}

\begin{table*}[!htb]
%\begingroup
%\squeezetable
%\begin{adjustbox}{width=1\textwidth}
\small
\caption{The DFT$+U$ total energy difference ($\Delta E$) after subtraction of two spin states. The $\Delta E$ values in the Table below are calculated using DFT$+U$ relaxed crystal structures with vdW ($\Delta E_\textrm{structure}^\textrm{from DFT$+U$+vdW}$) and without vdW ($\Delta E_\textrm{structure}^\textrm{from DFT$+U$}$) dispersion correction. The values of $n_i$ and $J_i$ are determined by the specific $\Delta E$ combination.}
\label{tab:J_cal}
\centering
\begin{tabular}{|c|c|c|c|}
\hline
$\Delta E$             & $\sum_{i} n_iJ_i$                 & $\Delta E_\textrm{structure}^\textrm{from DFT$+U$}$ (meV) & $\Delta E_\textrm{structure}^\textrm{from DFT$+U$+vdW}$ (meV) \\ \hline
$E_{FM} - E_{AFM1}$   & $8J_4$                            & 10.88              & 16.88              \\ \hline
$E_{FM} - E_{AFM2}$   & $4J_1 + 4J_3 +4J_4 +4J_5$         & 18.67              & 18.85               \\ \hline
$E_{FM} - E_{AFM5}$   & $8J_3 + 8J_5$                     & -1.95              & -3.08               \\ \hline
$E_{AFM3} - E_{AFM2}$ & $2J_1 - 4J_2 + 2J_3 - 2J_4 -2J_5$ & 12.36            & 9.59          \\ \hline
$E_{AFM1} - E_{AFM2}$ & $4J_1 + 4J_3 - 4J_4 +4J_5$        & 7.78              & 1.96               \\ \hline
$E_{AFM3} - E_{AFM4}$ & $4J_3 - 4J_4 - 4J_5$              & -6.63             & -10.25                   \\ \hline
\end{tabular}
%\end{adjustbox}
\end{table*}

\subsection{Selection of crystal structure for $J_n$ calculation}
Four different collinear spin configurations were explored, as shown in Fig.~\ref{fig:select_crystal}. The DFT$+U$ calculations predicted that the configuration shown in Fig.~\ref{fig:select_crystal}(b) has the lowest total energy. Thus, we used the relaxed crystal structure of this configuration to calculate the spin-exchange parameters.

\subsection{Computation of $J_n$}
\label{sec:calculateJ}
\begin{figure*}[!htb]
\begin{center}
\includegraphics[width=12.6cm]
{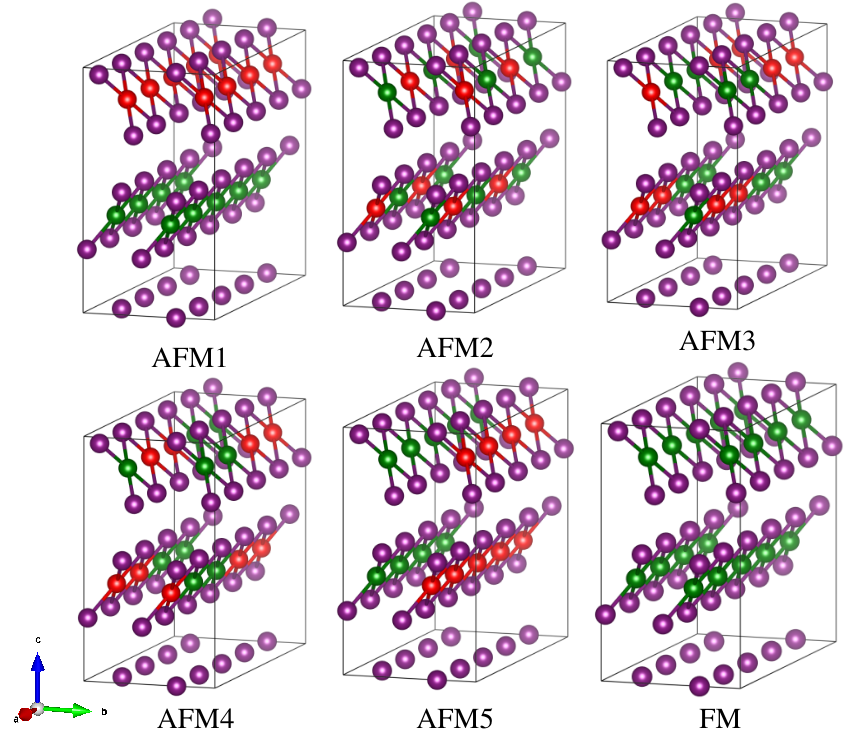}
\end{center}
\caption{Six ordered spin arrangement of CrI$_2$ generated to extract the spin-exchange parameter $J_1$ to $J_5$. 
FM refers to ferromagnetic, and AFM1 to AFM5 refer to five different anti-ferromagnetic configurations.
Spin-up and spin-down Cr atoms are represented in green and red colors, respectively. Iodine atoms are represented in purple color. The CIF files with the atomic parameters 
used to generate these supercells are included in the Supplementary Information.}
\label{fig:J_configures}
\end{figure*}

Six ordered spin arrangement of CrI$_2$ (shown in Fig.\ \ref{fig:J_configures}) were generated to extract the spin-exchange parameter from DFT calculations. The total spin-exchange energies for each ordered spin state for the supercell can be written as: 

\begin{align*}
    E_{FM} &= -8\left( 2J_1 + 2J_2 + 4J_3 + 4J_4 + 4J_5 \right)  \\
    E_{AFM1} &= -8\left( 2J_1 + 2J_2 + 4J_3 - 4J_4 + 4J_5 \right) \\
    E_{AFM2} &= -8\left( -2J_1 + 2J_2 + 0J_3 + 0J_4 + 0J_5 \right) \\
    E_{AFM3} &= -8\left( 0J_1 - 2J_2 + 2J_3 - 2J_4 - 2J_5 \right) \\
    E_{AFM4} &= -8\left( 0J_1 - 2J_2 - 2J_3 + 2J_4 + 2J_5 \right) \\
    E_{AFM5} &= -8\left( 2J_1 + 2J_2 - 4J_3 + 4J_4 - 4J_5 \right)
\end{align*}

\end{document}